\documentclass[]{article}

\usepackage{soul}
\usepackage{amsmath}
\usepackage{color,xcolor}
\usepackage{graphicx}
\usepackage{authblk}
\usepackage{hyperref}
\usepackage{cite}

\title{Single-Light-Pulse Driven Compact Atom Interferometry with Measurement Induced Large Momentum Transfer}

\author[1]{Yinghang Jiang}
\author[1]{Jiguo Wu}
\author[1]{Junfan Zhu}
\author[1]{Rongchun Ge \thanks{ Email: \href{rcge@scu.edu.cn}{rcge@scu.edu.cn}}}
\author[1]{Zhiyou Zhang \thanks{Corresponding author. Email: \href{zhangzhiyou@scu.edu.cn}{zhangzhiyou@scu.edu.cn}}}
\affil[1]{College of Physics, Sichuan University, Chengdu 610064, China}

\begin{document}
\maketitle

\begin{abstract}
	We propose a fundamentally new design strategy of light-pulsed atom interferometry (LPAI) with a single atomic beam splitter. A traditional $\pi/2$-pulse Raman beam is employed to render a small momentum transfer at the initial state. After a short period of evolution during which physical relevant information can be loaded, a quantum weak measurement is applied to the internal state of the atoms. The final information will be detected from the transmission spectrum of a probe light to obviate the measurement of florescence signal. An effective amplification of the order of $10^3$ about the momentum offset is achieved in our simulation employing $Cs$ atoms with current experimental condition. Our proposal offers a cost-effective, high-accuracy measurement and readout strategy for LPAI. Furthermore, the strategy makes the physical setup much simpler and more compact offering new direction towards portable sensitive LPAI. 
\end{abstract}

\section*{Introduction}
Coherence is at the heart of quantum theory or wavelike behavior in general. It is the ability that the system/object can be in different states simultaneously (with well defined phase relation once the basis are selected). These ubiquitous property of microscopic objects is most well demonstrated in phenomena of interference. However, for interference to happen ideally, the system need to be informationally isolated from the outside world since the unavoidable interactions will gradually spoil its coherence. Compared to its optical counterpart which has been not only the foundation of numerous modern technologies but also constantly driving forward our comprehension of the natural world, matter-wave interferometry was almost in its infancy until the later development of Laser technology and nanofabrication.

The possibility of an atom interferometry was investigated as early as the 1970s~\cite{RevModPhys.81.1051}. The versatility of electromagnetic response, inertial mass, and availability over different atomic species to just name a few have made it unique advantage for interferometry with atoms compared to electrons and neutrons. Light-puls e atom interferometry (LPAI) among others have been investigated intensive as a focus of great interest with potential and application for precision measurement such as gravity and gravitational acceleration \cite{nyman2006ice,le2008limits,Zhou_2011,PhysRevLett.97.060402,RevModPhys.81.1051,PhysRevA.86.043630}, gravitational constant \cite{doi:10.1126/science.1135459,PhysRevLett.100.050801,rosi2014precision}, fine structure constant \cite{weiss1994precision,PhysRevLett.101.230801}, gravitational wave \cite{PhysRevLett.110.171102,XUEFEI2015411}. Inspired by optical interferometry, a typical LPAI has three main components: beam spliter to split the source, mirror to reflect the atoms, and beam splitter again to get the interference~\cite{PLA89}. The first beam splitter is used to get different 'classical' paths by imparting a state dependent momentum to the atoms which serve as the histories like in Feynman's path integral. This is achieved by a $\pi/2$-pulse (stimulated Raman transition) in a traditional LPAI, but different strategies can be employed such as diffraction grating~\cite{RevModPhys.81.1051}, shaking the carrier lattice~\cite{Shakenlattice17,Shakenlattice}. The sensitivity of detection was shown to be proportional to the area surrounded by the LPAI~\cite{Path94}. So there is substantial endeavor to systematically pursue large momentum transfer atom optics recently~\cite{Multiphoton24,Bragg23,Floquet22,Swept15,Bloch09}. In the pursuit of large momentum transfer between the optical fields and the atoms in LPAI, the paths will travel away from each other; consequently, it seems mirrors are necessary to converge the atoms in real space to make it possible for different histories talk to each other. At the core of interference, it is the requirement of indistinguishability -- namely there should be non-zero contribution which you can not tell from which historical path it comes. Formally, when the above histories meet in space they need to have nonzero overlap for interference to happen. As in the typical LPAI, before the last beam splitter the internal states of different path are orthogonal to each other, the beam splitter is used to create a nonzero overlap for the internal states. At last the fluorescence signal of the population in excited energy levels are detected.

In this Letter, we propose a fundamentally new strategy towards compact LPAI driven by a single stimulated Raman-transition. Compared to the common wisdom of LPAI for which a large momentum transfer is pursued at the first beam splitter, we employ the well-developed stimulated Raman-transition to confer a momentum of $2\hbar k$ to the excited energy level $|e\rangle$ with respect to the ground state $|g\rangle$ of the atoms. Such a momentum difference is supposed to be much less than the bandwidth of the natural broadening of momentum distribution. Next follows a short time of evolution during which state dependent interactions can be loaded into the relative phase among different paths. At the same time, the spatial displacement between the wave packets are small compared to its natural spatial distribution which obviates the need of mirrors in the original setup. After that an operation named weak measurement which is best-known for its capable of boosting the detection of weak signals~\cite{PhysRevLett.60.1351,PhysRevX.4.011031,PhysRevLett.116.100803,denkmayr2014observation,PhysRevLett.126.100403,PhysRevLett.106.080405,article,PhysRevLett.102.020404,PhysRevLett.102.173601,PhysRevA.82.063822,Viza:13,PhysRevLett.109.013901,doi:10.1126/science.1152697,lundeen2011direct,mitchell2004super} is exerted on the atoms. Consequently, the atoms are effectively deposited into the ground state. Finally, the information of momentum distribution is retrieved from the transmission spectrum of a probe light instead of traditional method of fluorescence detection, and a momentum offset larger than $3$-orders of $\hbar k$ is observed.

\begin{figure}
\centering
	\includegraphics[scale=0.25]{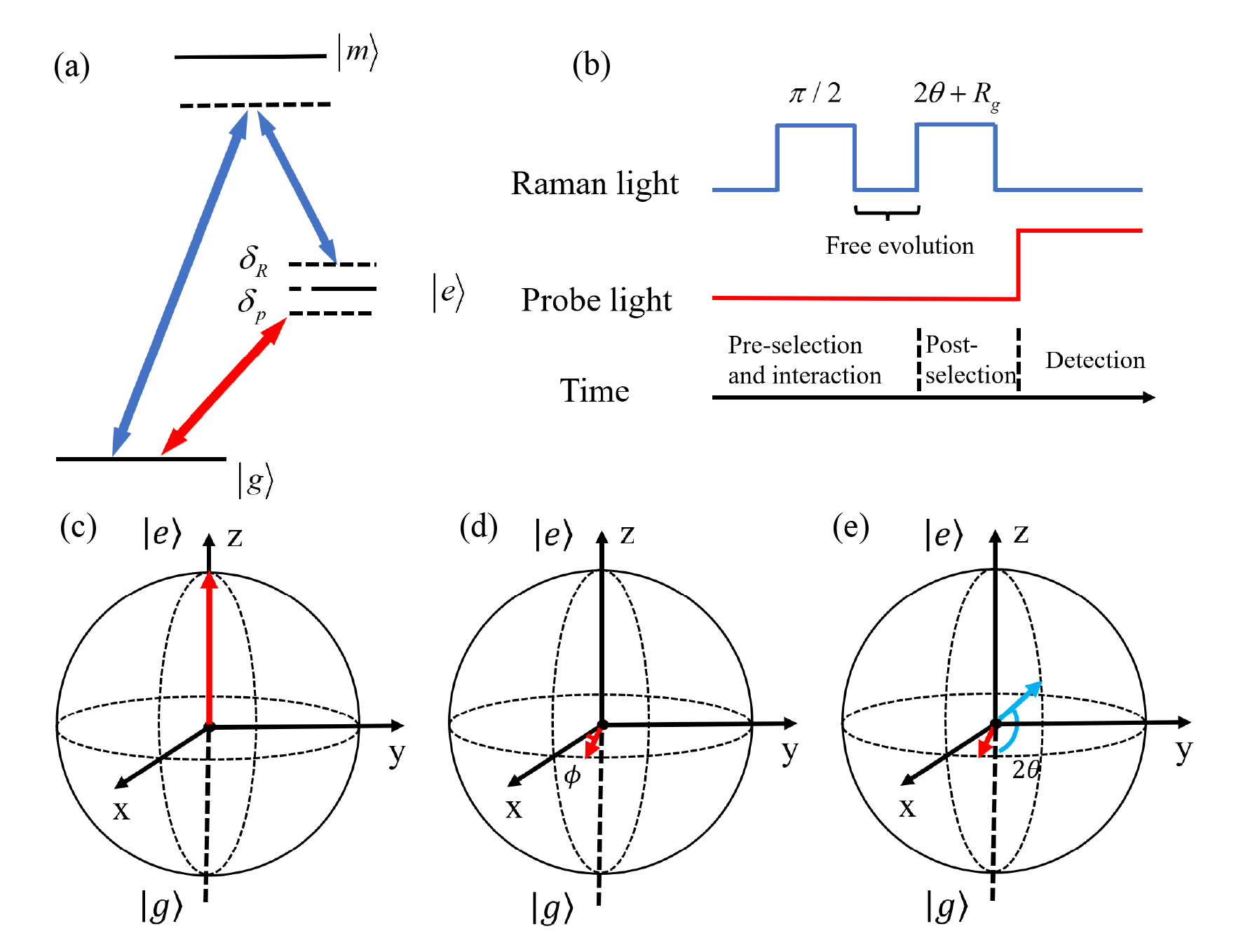}
	\caption{Experiment scheme. (a) Energy level diagram of experiment scheme. (b) Laser timing diagram of Ramam light and probe light. (c)-(e) Evolution of state vector in Bloch sphere, the red arrow represents the state vector, and the blue arrow represents the post-selection state.}
	\label{setup}
\end{figure}

\section*{Model and theory}
For the traditional three-level Raman-transition model of LPAI as shown in Fig.~\ref{setup}(a), where two (meta) stable energy levels ($|g\rangle$, $|e\rangle$), are loosely coupled to a third transient channel $|m\rangle$ with Rabi frequencies $\Omega_1$ and $\Omega_2$ by two laser beams $\omega_1$ and $\omega_2$ respectively. When the condition of large detuning, $\omega_m-\omega_1-\omega_g\equiv \Delta\sim \omega_m-\omega_2-\omega_e\gg |\Omega_{1/2}|$ (for near resonant two-photon process), is satisfied, the transient level can be adiabatically eliminated as (up to order $O(|\Omega/\Delta|)$),
\begin{equation}
	\begin{aligned}
		H=&\frac{\widehat{\mathbf{p}}^2}{2 m}+\hbar \frac{\omega_a}{2}\hat{\sigma}_z-\left[\hbar \Omega e^{i\mathbf{k}\cdot \hat{\mathbf{r}}}\hat{\sigma}_++h.c.\right].
		\label{eq1}
	\end{aligned}
\end{equation}
Here the first term is the kinetic energy of the atom; $\hat{\sigma}_z = |e\rangle\langle e|-|g\rangle\langle e|$, $\hat{\sigma}_+$ are the third Pauli matrix and the raising operator. $\Omega = \frac{\Omega_1\Omega_2^*}{2\Delta}$ is the effective Rabi frequency between the simplified two-level system; $\omega_a = (|\Omega_1|^2-|\Omega_2|^2)/\Delta + \omega_1+\omega_g-\omega_2-\omega_e$, and ${\bf k} = {\bf k}_1-{\bf k}_2$ where we assume counter-traveling laser beams as usual. In the following analysis, we will take $|\Omega_1| = |\Omega_2|$ for simplicity.

For atoms prepared in ground state $|g\rangle$ with momentum distribution $|\phi({\bf p})\rangle \equiv \int \varphi({\bf p}+{\bf p}')|{\bf p}+{\bf p}'\rangle d{\bf p}'$, after a $\pi/2$-pulse, the system is partially excited. Depending on the internal state, there is a offset of momentum of $\hbar {\bf k}$ in the distribution $|g,{\bf p}\rangle\leftrightarrow |e,{\bf p+\hbar k}\rangle$. This momentum dressing will regularize the correspondent energy splitting as $\omega_a\rightarrow \delta_R = \omega_a + \frac{(\bf p+\hbar k)^2 - {\bf p}^2}{2m\hbar} = \omega_a + v_k |{\bf k}| + \hbar{\bf k}^2/2m$\cite{kasevich1992measurement,cadoret2009atom}. For later convenience, we will make a frame transformation from the lab frame to moving frame with velocity $\frac{\hbar{\bf k}}{2m}$, then the effect of the pulse can be mathematically represented as $\hat{U}(\frac{\pi}{2}) = e^{\frac{i}{2}\hat{\sigma}_z\hbar{\bf k}\cdot\hat{\bf r}}e^{-i\frac{\pi}{4}\hat{\sigma}_x}$ (assuming the external motion of the atoms is negligible during the excitation process), which takes the atoms from $|g\rangle|\phi({\bf p})\rangle$ to $1/\sqrt{2}\Big(|g\rangle|\phi({\bf p}-\hbar{\bf k}/2)\rangle-i|e\rangle|\phi({\bf p}+\hbar{\bf k}/2\rangle\Big)$. Physically, it induces a momentum shift of $\hbar{\bf k}$ for the excited state $|e\rangle$ compared to $|g\rangle$.

However, the splitting of the two momentum wave packets above is too small. In order to achieve a high sensitive interferometer, there have been substantial experimental interests to achieve large momentum transfer atom optics. In stead, in the following we propose a new strategy employing the technique of weak measurement to simplify the physical setup of LPAI, which takes the full advantage small of splitting induced by the stimulated Raman-transition and show an amplification of momentum transfer as large as $10^3$ times of $\hbar {\bf k}$. To be more specific, consider a group of cold atoms with small velocity (distribution) $|{\bf v}_g\rangle$ in ground state $|g\rangle$. As is explained before there are two component of the evolution, the $\pi/2$-pulse makes it $|g\rangle\rightarrow |\psi_{\frac{\pi}{2}}\rangle = 1/\sqrt{2}\Big(|g\rangle - i|e\rangle\Big)$; at the same time,  with a $\pi/2$-pulse, both the internal state and momentum wave are changed being entangled with other, which plays the role of weak interaction from the perspective of weak measurement. We call it ``weak'' since the typical momentum shift resulted is around $10^{-25}$~g$\cdot$m/s, which is much smaller than natural broadening in momentum space. Then we have the whole state including momentum wave function $|{\bf v_g}\rangle$ given as $|\Psi\rangle = \exp(i\frac{\hat{\sigma}_z}{2}\hbar{\bf k}\cdot\hat{\bf r})|\psi_{\frac{\pi}{2}}\rangle \otimes|{\bf v_g}\rangle$, where $\hat{\sigma}_z$ is the Pauli operator, $\hat{\bf r}$ is the position operator. In a word, the pulse partially pumps the atoms into $|e\rangle$ with velocity ${\bf v_e} = {\bf v}_g + \hbar {\bf k}/2m$. Then after the free evolution with period $T_f$ and probably with local fields for which the interaction is diagonal in the basis of $|g\rangle,|e\rangle$ for which the total effect is given by $U_{T_f}$, the whole system is now being:
 
\begin{equation}
	\begin{aligned}
		|\Psi_f\rangle &= U_{T_f}\exp(i\frac{\hat{\sigma}_z}{2}\hbar {\bf k} \cdot\hat{\bf r})|\psi_{\frac{\pi}{2}}\rangle|{\bf v_g}\rangle \\
		 &\approx \exp(i\frac{\hat{\sigma}_z}{2}\hbar {\bf k} \cdot\hat{\bf r}) U_{T_f}|\psi_{\frac{\pi}{2}}\rangle|{\bf v_g}\rangle \\
		 &= \exp(i\frac{\hat{\sigma}_z}{2}\hbar {\bf k} \cdot\hat{\bf r})\Big(|g\rangle e^{i\phi_g}-i|e\rangle e^{i\phi_e} e^{i\phi_f}\Big)|{\bf v_g}\rangle /\sqrt{2}.
	\end{aligned}
\end{equation} 

Where $\phi_g=\frac{1}{2}m{\bf v}_g^2\times \frac{T_f}{\hbar}$, $\phi_e=\frac{1}{2}m{\bf v}_e^2\times \frac{T_f}{\hbar}$ are the phases acquired through free evolution, $\phi_f$ is the local field induced relative phase, and we assume both the result of noncommutative (proportional to time).

\begin{figure}
\centering
\includegraphics[scale=0.35]{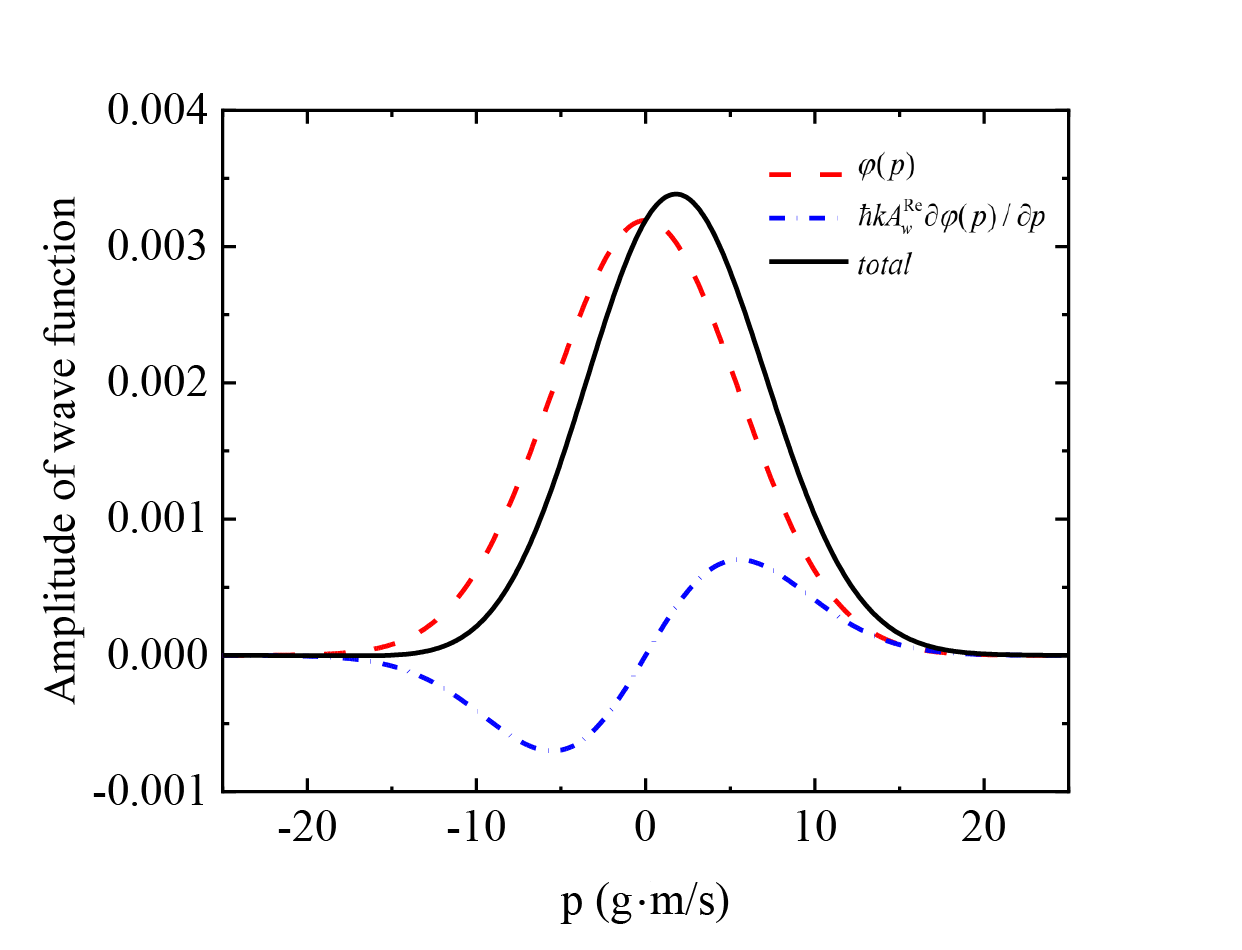}
\caption{Asymmetric interference of momentum wave packets of cesium atoms at $1\mathrm{K}$. The symmetric (red), asymmetric (blue) parts and total (black) of the post-selection. The centroid of the wave packet is significantly shifted. ($\theta=\pi /4-\pi /1000$ and $\phi=0.03$) } 
\label{coherence}
\end{figure}

For a short period of evolution after the $\pi/2$-pulse, the spatial displacement between the wave packets corresponding to different internal states ($|g\rangle$, $|e\rangle$) is much smaller than spatial distribution of the initial wave packet, as a result the different `classical' paths are spatially overlapped. This obviates the need of mirror in the traditional LPAI. Then a weak measurement is carried out with a post-selection which projects the internal state of the system to $|\psi_p\rangle=\cos\theta|g\rangle  +i\sin\theta |e\rangle$. What we should clarify that post-selection itself is a kind of projection measurement (strong measurement), and it is part of the so-called weak measurement mentioned earlier. Consequently, the conditional momentum distribution (with initial distribution $|\phi({\bf p})\rangle$ of the atoms are now given as (to the first order of $\hbar{\bf k}$),
\begin{equation}
   \begin{aligned}
	&\langle\psi_p|\exp(i\frac{\hat\sigma_z}{2}\hbar{\bf k}\cdot\hat{\bf r})|\psi_i\rangle|\phi({\bf p})\rangle\approx \langle\psi_p|1-i\hat{A}k\hat{x}|\psi_i\rangle |\phi({\bf p})\rangle \\
	=&\langle\psi_p|\psi_i\rangle [1+\hbar k A_w^{Re}\frac{\partial}{\partial p}-i\hbar k A_w^{Im} \frac{\partial}{\partial p}] |\varphi_p\rangle.
	\label{eq8}
   \end{aligned}
\end{equation}

Here for simplicity we have used $|\psi_i\rangle \equiv \Big(|g\rangle -i e^{i\phi}|e\rangle\Big)/\sqrt{2}$, $\phi=\phi_e-\phi_g+\phi(t)$, $\hat{A} = \hbar\frac{\hat{\sigma}_z}{2}$, and the laser beams are assumed in $x$-direction. In Eq.\ref{eq8}, $A_w^{Re}=Re(\frac{\langle \psi_p|\hat{A}|\psi_i\rangle}{\langle \psi_p|\psi_i\rangle})=1+\frac{2\cos\phi \cot\theta +2}{\cot^2\theta-2\cos\phi\cot\theta+1}$, $A_w^{Im}=Im(\frac{\langle \psi_p|\hat{A}|\psi_i\rangle}{\langle \psi_p|\psi_i\rangle})=\frac{-2i\sin\phi \cot\theta}{\cot^2\theta-2\cos\phi\cot\theta+1}$ and $\hat{x}=-i\hbar\partial/\partial p$. 
It can be shown that as a result the post-selection the real part of weak value $A_w^{Re}$ causes the wave functions of the two orthogonal states ($|\psi_i\rangle$ and $|\psi_{i\perp}\rangle$) to interfere in the momentum space, while the imaginary part of weak value does not because it is out of phase a $\pi/2$ phase difference. The momentum offset caused by this asymmetric interference effect can easily reach the FWHM level of the momentum wave packet even with an input of momentum from the Raman-transition which is several orders smaller. As is shown clearly in Fig.~\ref{coherence}, a substantial displacement of the centroid of the momentum distribution is observed.   
The post-selection wave packet amplitude is the total of these two parts: symmetric $\varphi({\bf p})$ (red dashed) and asymmetric given by the second term of of Eq.~(\ref{eq8}) (blue dashed). Consequently, the larger the real weak value is, the larger the contribution of the asymmetric part is, and the offset of the centroid is significant. 

The above-mentioned process seems difficult to understand. Why would the position of the momentum wave packet shift and be magnified several times in the absence of momentum transfer? An intuitive explanation is offered to understand the process. The atomic momentum wave packet and the atomic internal state are entangled through the first $\pi /2$ pulse. Post-selection is an operation on the internal state of atoms, but in fact, it is also an operation on the momentum wave packets, because they are already entangled together. At the same time, Post-selection retains a portion of atoms with large momentum shifts while screening out the majority. Therefore, we can eventually observe that the momentum is amplified. 

The working temperature of the interferometry is also a matter worthy of discussion. Traditional interferometry requires the complete separation of the momentum wave packets to achieve higher sensitivity, which is easily realized at low temperatures. In contrast, our scheme dependent on the interference between wave packets, requiring the packets to be slightly separated. Therefore, the latter only needs a very short period of evolution time and is suitable for use in hot temperatures. Although the thermal motion of the atoms is intense and leads to collision decoherence in hot environment, evolution time at sub-microsecond level could preserve coherence to the greatest extent. The scheme holds the potential to achieve the sensitivity of a cold atomic interferometry in a hot environment.

In order to achieve the proposed weak measurement of the internal states of atoms in experiment, additional control pulse will be needed to achieve an effective transformation of $|\psi_p\rangle\rightarrow |g\rangle$. With the inspiration of quantum computation, this amounts to find the quantum operation/gates to achieve
\begin{equation}
\textbf{R}_{post} = \textbf{U}_{2\theta}\textbf{U}_{-2\theta}\textbf{R}_{post }= \textbf{U}_{2\theta}\textbf{R}_g ,
\label{eq12}
\end{equation}
where $\textbf{R}_{post}$ and $\textbf{R}_{g}$ represent projection operation on the post-selection state and the ground state, $\textbf{U}_{2\theta}$ and $\textbf{U}_{-2\theta}$ represents action of $2\theta$ and $-2\theta$ light pulse (rotate $2\theta$ and $-2\theta$ around the x axis on the Bloch sphere). This sequential operations of laser-pulse is shown in Fig.~\ref{setup}(b).
The atoms are initially polarized to an excited state, located directly above the Bloch sphere in Fig. \ref{setup}(c). The first Rabi light pulse of $\pi/2$ in $y$ direction immediately evolves the state vector into the $xoy$ plane. After a period of free evolution $\tau$, the state vector acquires the kinetic phase and get the angle $\phi$ with $x$ axis in Fig. \ref{setup}(d). Then, strong measurement to post-selection $|\psi_f\rangle$ is achieved via a $2\theta$ pulses around the $y$-axis and state-dependent fluorescence. Finally, the probe light is incident along the pump light direction, and the probe light frequency is scanned to readout the centroid of spectrum. 

\section*{Results and discussion}
In the following, we will present more practical analysis with in the reach of current experimental setup.
Using $Cs$ atoms at $1~\mathrm{K}$ as an example. The de Broglie thermal wavelength of cesium atoms ($10^{-12}\mathrm{m}$) is much smaller than their mean free path ($10^{-7}\mathrm{m}$). So they are far away from quantum degeneracy \cite{PhysRevLett.74.3352,PhysRevLett.75.3969,cornell1996very,london1938lambda}, and can be well represented single atoms without non-local quantum correlations. We will model that statistics of atom employing the 
classical Maxwell-Boltzmann distribution,
\begin{equation}
\varphi(p)=(\frac{m }{2\pi k_B T})^{1/2} e^{\frac{-p^2}{2k_B Tm}} ,
\label{eq9} 
\end{equation}
where $m$ is mass of atom, $k_B$ is Boltzmann constant, $T$ is thermodynamic temperature. It describes the initial momentum distribution prior to evolution or measurement.

\begin{figure}
\centering
\includegraphics[scale=0.3]{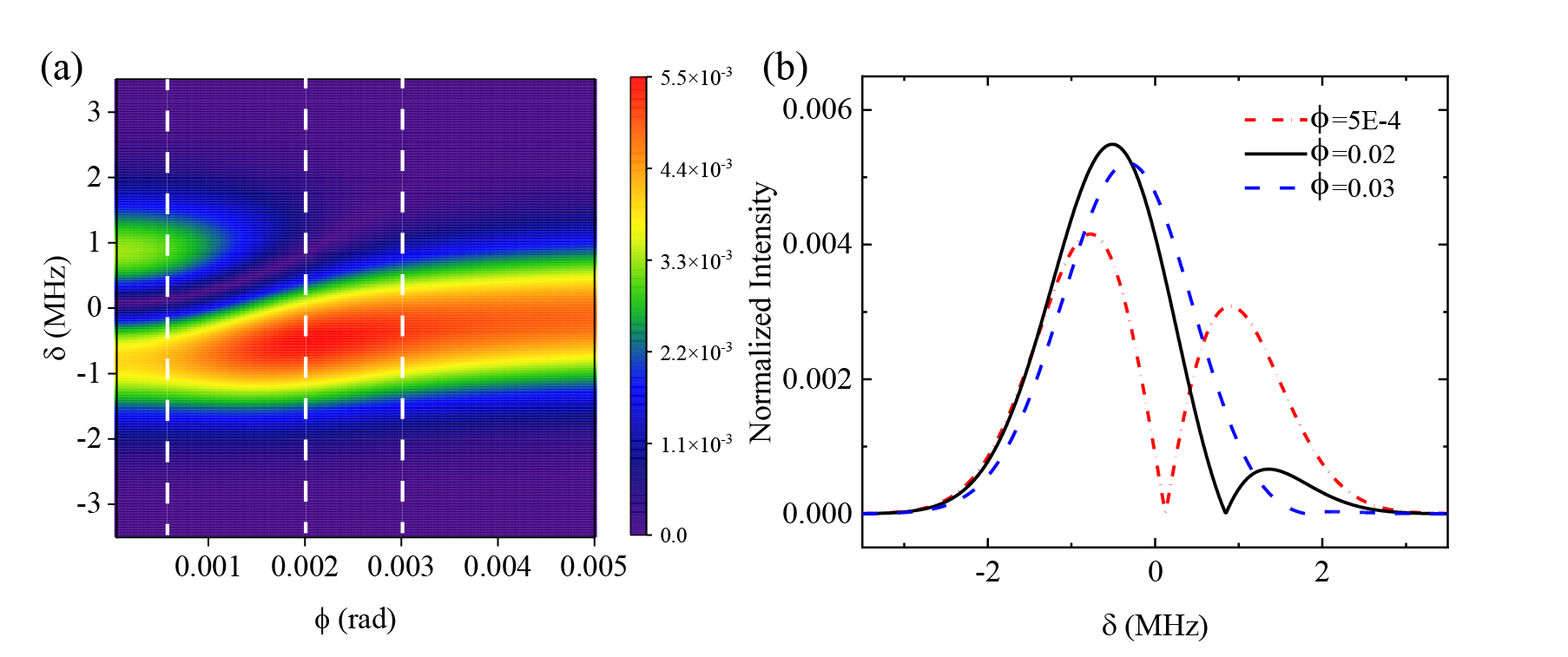}
\caption{Transmission spectrum of probe light considering the Doppler effect. (a) Normalized heat map of Doppler effect with laser frequency detuning $\delta$ and post-selection parameters $\phi$. (b) Transmission spectrum with different post-selection parameters, red dotted line for $\phi=0.0005$, black solid line for $\phi=0.02$, and blue dotted line for $\phi=0.03$. The temperature $T=1K$ and $\theta=\pi /4-\pi /1000$. }
\label{pic1}
\end{figure}

The momentum distribution of the atoms after the weak measurement above can be achieved by inspecting the transmission spectrum of a probe light exploiting the Doppler sensitivity of atoms. For simplicity, we consider the atoms moving along the $x$ direction. The momentum wave packet after the post-selection as a function of the velocity from Eq. (\ref{eq8}) and Eq. (\ref{eq9}) is 
\begin{equation}
F(v)=A_0[f(v)-\hbar k A_w^{Re}\frac{v}{K_BT}f(v)],
\label{eq10}
\end{equation}
where $A_0$ is a constant for normalization, $f(v)=(\frac{m }{2\pi K_BT})^{1/2}\exp(-\frac{mv^2}{2K_BT})$ is the Maxwell-Boltzmann velocity distribution. Considering the Doppler effect, the optical frequency and the atomic velocity have the following relationship, $w=w_0(1+v/c)$ and $dv=(c/w_0)dw$, where $w_0$ is the eigen-frequency of atoms. Absorbed power scales with the atomic density within the frequency interval $dw$ (as is shown later that weak measurement will effectively project the atoms into ground state), thus the intensity profile of Doppler broadening is obtained as
\begin{equation}
 \begin{aligned}
	I(w)=&|I_0\{exp[-(\frac{c(w-w_0)}{w_0 V})^2]-\hbar k A_w^{Re}(\frac{c(w-w_0)}{w_0}) \\
	&\frac{1}{K_BT}exp[-(\frac{c(w-w_0)}{w_0 V})^2]\}|,
 \end{aligned}  
\end{equation}
where $V=(2K_BT/m)^{1/2}$. We simulated the Doppler effect of probe light which couples ground energy level $6S_{1/2}$ and excited energy level $8P_{3/2}$ at a temperature of $1K$. The heat map of the spectrum intensity are shown in Fig. \ref{pic1}(a), and spectrum of $\phi=0.0005,0.02$ and $0.03$ are shown in Fig. \ref{pic1}(b). When the parameter $\phi$ is less than $0.02$, the spectral lines show obvious bimodal characteristics and have significant centroid offset. 

\begin{figure*}
\centering
\includegraphics[scale=0.3]{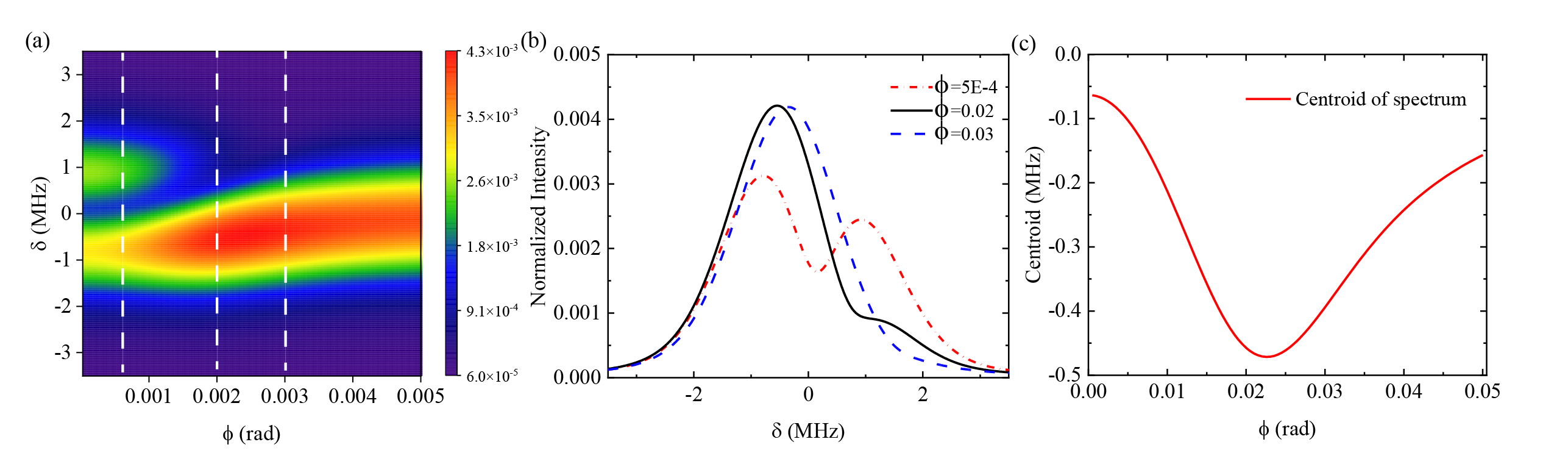}
\caption{Results of simulation. (a) Normalized heat map of transmission spectrum of probe light with laser frequency detuning $\delta$ and post-selection parameters $\phi$. (b) Transmission spectrum with different post-selection parameters, red dotted line for $\phi=0.0005$, black solid line for $\phi=0.02$, and blue dotted line for $\phi=0.03$. (c) Relationship between spectral centroid offset (compared to frequency detuning $\delta$) and $\phi$. The temperature $T=1K$ and $\theta=\pi /4-\pi /1000$.}
\label{pic3}
\end{figure*}

In realistic situation, the spectrum is further affected by other broadening mechanisms, and the most important of which is the natural broadening caused by the spontaneous emission of excited state atoms. The natural linewidth of a Se atom at $8P_{3/2}$ is about $0.53\mathrm{MHz}$, which is comparable with doppler broadening. Considering natural broadening, not all atoms with a definite velocity component $v_0$ can emit or absorb photons at $w=w_0(1+v_0/c)$. The final spectral line can be obtained by convolution with Lorenz line pattern from natural broadening. When the natural broadening is greater than twice the broadening caused by the Doppler effect, the natural linewidth will cover the information in the spectral line. Therefore, we should keep the Doppler broadening greater than $0.53$ MHz and this also requires adequate temperature. Another broadening mechanism that is often considered in atomic gases is the transit time broadening, which comes from the fact that the interaction time of the atoms and light field is less than the lifetime of the excited state. In our case, the lifetime of the excited atoms is on the order of microseconds, while it takes about $3\mathrm{ms}$ for the cesium atoms to pass through a $1\mathrm{mm}$ laser beam at $1K$ temperature. Therefore, the transit time broadening does not need to be considered.

With the above consideration, we simulate the heat map of spectrum intensity including natural broadening. As expected, the spectra are significantly broadened compare to the case of Fig.~\ref{pic1}, but the moving trend of centroid is consistent as previous. We plot the spectral centroid as a function of the post-selection parameter $\phi$ as shown in Fig.\ref{pic3}. When $\phi$ is less than $0.02\ rad$, the centroid of the spectrum still shows a strong linear dependence on the parameter, while $\phi$ is greater than $0.02\ rad$, the trend is opposite and the slope is gradually slowing down. Obviously, as $\phi$ increases, the centroid tends to the original position, because the effect of post-selection is gradually weakened. Besides, it can be seen from Fig.\ref{pic3}(c) that the centroid offset is at sub-MHZ level, which has a great advantage over the fluorescence readout of traditional atom interferometer. Latter relies on charge coupled devices and stray photons in the environment greatly affect the measurement accuracy. This results in high experimental cost, high environmental requirements and low accuracy of conventional atomic interferometers. In contrast, asymmetric interferometers use the detection spectrum to read out the signal, the current spectral measurement resolution can be easily achieved level of sub-Hz at low cost, and lower environmental requirements.

Finally, we analyzed the phase sensitivity characteristics of the Single-Light-Pulse driven atom interferometry and common atom interferometry. For a classic $\pi/2 -\pi/2$ atom interferometry, the probability to find the atom in the excited state is $\frac{1}{2} (1+\cos{\phi})$\cite{cadoret2009atom}. This is a standard form of an interference pattern, the atomic fluorescence intensity in the excited state and the kinetic phase have a trigonometric function relationship. This phase term consists of three parts, the first two depend on kinetic energy: $\phi_{g/e}=\frac{1}{2} m \boldmath{\nu}_{g/e}^2 \times \frac{T}{\hbar}$, the last term depends both on the laser frequency and on the internal energy of the atoms. Under condition of zero detune: $\phi=(\frac{\boldmath{\nu}_e+\boldmath{\nu}_g}{2})(\boldmath{\nu}_e-\boldmath{\nu}_g)\frac{mT}{\hbar}$. The phase $\Phi$ depends on mean velocity of atoms, thus it can be used to probe the initial velocity distribution and coherence properties of atom gas \cite{hugbart2005coherence}. Standard quantum limit (SQL) of atom interferometry is :$\delta\Phi=\frac{1}{\sqrt{N}}$ \cite{kasevich1991atomic}. However, due to decoherence caused by environmental noise and other factors, the ideal sensitivity cannot be achieved in the experiment of cold atom interferometry, while hot atom interferometry is much lower in comparison because of atomic collision decoherence and high phase noise. 

\begin{figure}
\centering
\includegraphics[scale=0.28]{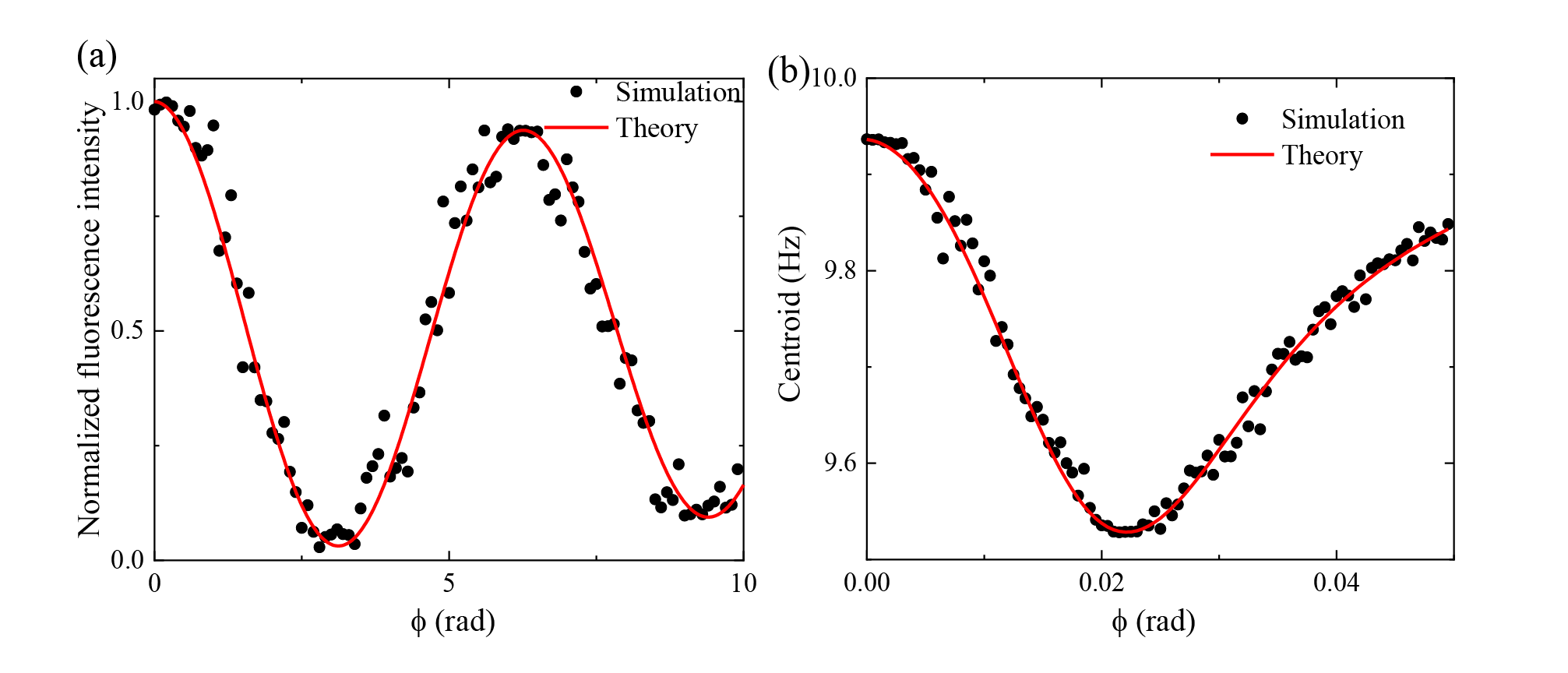}
\caption{Results of simulation. (a) Interference fringes of traditional atom interferometry adding white noise. (b) Centroid of spectrum of compact LPAI adding white noise. The signal-to-noise ratios of the two simulation scenarios are the same.}
\label{noise}
\end{figure}

We conducted a simulated comparative experiment between the traditional interferometry and the new scheme. Just as we mentioned earlier, decoherence disrupts the phase correlation between the two internal states, therefore, random white noise is added to the phase signal to simulate the effect of decoherence. It can be seen from the simulation results in FIG. \ref{noise} that due to the weak value amplification effect, our scheme can respond to tiny phase shifts ($\textless 0.01rad$). A smaller measurable phase implies that a shorter evolution time is required, thereby effectively reducing a portion of the relaxation noise caused by decoherence. Therefore, our scheme can theoretically enhance the sensitivity of the atomic interferometry.

This atomic interferometry scheme exhibits significantly broader application potential compared to traditional approaches. Perturbations can be readily introduced during the free evolution time of the atoms, ultimately manifesting as a phase shift in the interference pattern. This phase difference is further amplified via weak measurement techniques, thereby enabling the extraction of information about the perturbation.

Two primary mechanisms can induce such perturbations and generate measurable phase differences. The first arises from electromagnetic interactions between the atomic magnetic or electric dipole moment and external fields. For example, consider the coupling of atoms with a far-detuned electric field characterized by a Rabi frequency $\Omega$. Given that the detuning $\delta$ satisfies $\delta\gg\Omega$, the electric field does not induce effective transitions between the ground state $|g\rangle$ and the excited state $|e\rangle$. Consequently, the populations in these states remain constant over time. However, the AC Stark effect \cite{bakos1977ac} induces energy shifts: $\Delta E_g=\frac{\hbar \Omega^2}{4\delta}$ for the ground state and $\Delta E_e=-\frac{\hbar \Omega^2}{4\delta}$ for the excited state. This results in an additional phase difference of $\frac{\Omega^2 \tau }{2\delta}$ accumulated over the free evolution time $\tau$. By controlling the evolution time and knowing the average atomic velocity, this phase shift can be precisely measured. Furthermore, the spectral pointer shift is amplified by the imaginary weak value according to: $\delta p \propto \frac{\Omega^2 \tau Im(A_w)}{2\delta}$. This effect can be utilized to measure electric field intensity or other parameters of external fields that interact with the atomic electromagnetic moments. The second mechanism occurs when the parameters of the system’s Hamiltonian undergo slow cyclic variations, causing the wave function to accumulate a geometric (Berry) phase dependent solely on the path traversed in parameter space. For a two-level system, this parameter space corresponds to the direction on the Bloch sphere. By modulating the phase and amplitude of an external field, the effective magnetic field direction can be made to trace a closed path, thereby generating a Berry phase. During the free evolution time, the interaction between atoms and an external electric field can be described in the rotating frame by the effective Hamiltonian: $H_{eff}=\frac{\hbar \delta}{2}\sigma_z+\frac{\hbar \Omega_z}{2}[\cos \phi(t)\sigma_x+\sin \phi(t)\sigma_y]$,
where $\phi(t)=\alpha t$ denotes the time-dependent phase of the electric field. Under the adiabatic condition, wherein $\phi(t)$ varies sufficiently slowly, the Berry phase is determined by the cyclic evolution of the effective magnetic field direction. This results in the accumulation of a geometric phase proportional to the solid angle $\Omega_s=2\pi(1-\cos\theta)$, where $\theta=\arctan(\Omega/\delta)$. The Berry phase is given by: $\gamma=\frac{-\Omega_s}{2}$, and it can similarly be amplified via the imaginary weak value: $\Delta p=Im(A_w)\gamma$. The Berry phase is highly sensitive to minute changes in system parameters, making it a powerful tool for precision sensing and metrology \cite{aidelsburger2013realization}. Moreover, it plays a fundamental role in various novel quantum phenomena—such as the topological properties \cite{atala2013direct} of atomic systems and the quantum Hall effect \cite{thouless1982quantized}—making its accurate measurement essential for validating theoretical models and advancing quantum technologies.

In conclusion, we have proposed a new design strategy of LPAI. In our proposal only a small initial momentum splitting of atoms is required, so a simple $\pi/2$ Raman-transition pulse is employed instead of the more demanding technique for large momentum transfer. As a result mirrors employing additional composite controlling codes are obviated in the physical setup. Then a weak measurement is applied to the internal states of atoms after a short period of evolution, which finally set the atoms in ground state. Consequently, the physically relevant information are deposited in the momentum distribution of atoms, which can be obtained the centroid offset of transmission spectrum. We showed that sub-MHZ level measurement is achieved by using the momentum wave packet of the atom as the measurement pointer. Our proposal does not only significantly simplify the current setup of LPAI, but also renders lower noise and higher measurement accuracy which will offer new opportunities for the application of LPAI.

\subsection*{Disclosures} The authors declare no conflicts of interest.
	
\subsection*{Data availability} Data underlying the results presented in this paper are not publicly available at this time but may be obtained from the authors upon reasonable request.

\bibliographystyle{unsrt} 
\bibliography{ref1.bib}
\end{document}